\documentclass[reqno,10pt]{article}

\usepackage[english]{babel}
\usepackage[utf8]{inputenc}
\usepackage[a4paper, total={6.5in, 8.6in}]{geometry}
\usepackage{tikz}

\usepackage{amsmath}
\usepackage{authblk}
\usepackage{graphicx}
\usepackage[hidelinks]{hyperref}
\usepackage{mathtools}
\usepackage{multicol}
\usepackage[normalem]{ulem}
\usepackage{xcolor}
\usepackage{amssymb}

\hypersetup{
  colorlinks   = true, 
  urlcolor     = blue, 
  linkcolor    = blue, 
  citecolor   = blue 
}

\newcommand{\bestbound}{0.8971}

\newcommand{\highestp}{17}
\newcommand{\gammav}{{\boldsymbol{\gamma}}}
\newcommand{\betav}{{\boldsymbol{\beta}}}
\newcommand{\paramv}{\gammav,\betav}
\newcommand{\approxgammav}{\boldsymbol{\tilde{\gamma}}}
\newcommand{\approxbetav}{\boldsymbol{\tilde{\beta}}}
\newcommand{\approxparamv}{\approxgammav,\approxbetav}
\newcommand{\costperedge}{c_{\rm edge}}

\DeclarePairedDelimiter{\floor}{\lfloor}{\rfloor}

\title{Lower bounding the MaxCut of high girth 3-regular  graphs  using the QAOA}

\author[1,2]{Edward Farhi}  
\author[ ]{Sam Gutmann}
\author[3]{Daniel Ranard}
\author[1]{Benjamin Villalonga}

\affil[1]{\textit{\footnotesize{Google Quantum AI, Venice, CA 90291}}}
\affil[2]{\textit{\footnotesize{Center for Theoretical Physics, Massachusetts Institute of Technology, Cambridge, MA 02139}}}
\affil[3]{\textit{\footnotesize{Walter Burke Institute for Theoretical Physics, California Institute of Technology, Pasadena, CA 91125}}}

\date{\today}

\begin{document}
\maketitle

\begin{abstract}
 We study MaxCut on 3-regular graphs of minimum girth $g$ for various $g$'s. We obtain new lower bounds on the maximum cut achievable in such graphs by analyzing the Quantum Approximate Optimization Algorithm (QAOA). For $g\geq16$, at depth $p \geq 7$, the QAOA improves on previously known lower  bounds. Our bounds are established through classical numerical analysis of the QAOA's expected performance. This analysis does not produce the actual cuts but establishes their existence.   When implemented on a quantum computer, the QAOA provides an efficient algorithm for finding such cuts, using a constant-depth quantum circuit.  To our knowledge, this gives an exponential speedup over the best known classical algorithm guaranteed to achieve cuts of this size on graphs of this girth. We also apply the QAOA to the Maximum Independent Set problem on the same class of graphs.
\end{abstract}

\section{Introduction}
The MaxCut problem asks one to partition the vertices of a graph into two subsets to maximize the size of the ``cut,'' or the set of edges that cross between the subsets. Determining the maximum cut of a graph is NP-hard, one of Karp's original 21 problems \cite{karp1975computational}, and it remains NP-hard when restricted to 3-regular graphs \cite{yannakakis1978node}. 
Often one studies MaxCut as an approximate optimization problem, with the goal of finding a large cut.

Here we study MaxCut on 3-regular graphs with large girth. The girth of a graph is the length of its shortest cycle.  We are interested in two questions: (1) How can we lower bound the maximum cut of a graph in terms of its girth, and (2) How can we use a classical or quantum algorithm to efficiently obtain a cut at least this large?
We answer both questions through our analysis of the Quantum Approximate Optimization Algorithm (QAOA) \cite{farhi2014quantum}.
While the first question is a mathematical question that does not concern algorithms, we prove a new lower bound constructively, by proving a performance guarantee on the QAOA applied to 3-regular graphs of sufficient girth. Here we have analyzed the performance of the QAOA without the need to actually run the algorithm. To find a cut of that size using the QAOA, you need to run the algorithm on quantum hardware. To our knowledge, the QAOA then provides an exponential speedup over the best known classical algorithms guaranteed to achieve a cut of this size, which require exponential time.

We quantify a cut by the cut fraction, that is,  the fraction of edges that appear in the cut.  For a graph $G$, we can ask about the cut fraction of the largest cut, or $\text{MaxCut}(G)/|E(G)|$ for edge set $E(G)$.
The smallest possible maximum cut fraction among all 3-regular graphs of girth at least $g$ is denoted
\begin{align} \label{eq:Mg}
    M_g = \inf_{\substack{\text{3-regular graphs }G\\\text{girth(G)}\geq g}} \frac{\text{MaxCut}(G)}{|E(G)|}.
\end{align}
We use the infimum because the set of graphs $G$ with girth($G$)$\geq g$ is infinite,  but \eqref{eq:Mg} means that every graph $G$ of girth at least $g$  has a cut of size at least $|E(G)| M_g$.  The numerical value of $M_g$ is currently unknown for general $g$.

\begin{figure}
    \centering
    \includegraphics[width=0.50\linewidth]{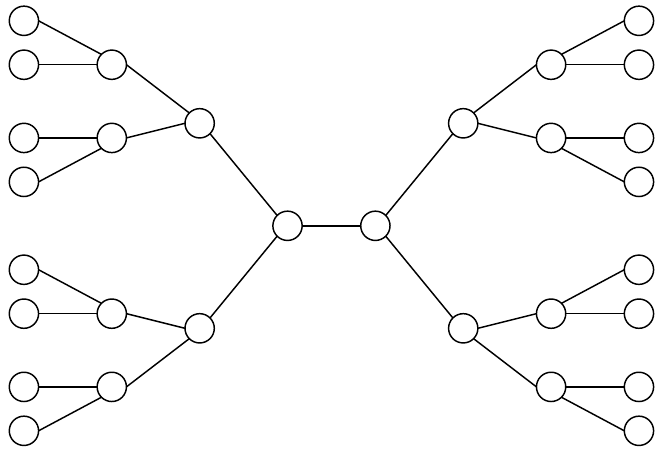}
    \caption{Edge neighborhood of the central edge of a 3-regular graph at $p=3$ and girth $g\geq8$. Every edge in the graph has this neighborhood structure at this minimum girth. We evaluate $\tilde{c}_{\rm edge}(p)$ which is the optimal quantum expected value of the cost function on the central edge.}
    \label{fig:regular_tree}
\end{figure}

In this paper, we obtain new rigorous lower bounds on $M_g$.   We do this by predicting the performance of the QAOA for MaxCut on large girth 3-regular graphs. MaxCut may be viewed as an optimization problem over bit strings of length $|V(G)|$, where $V(G)$ is the vertex set, and each bit string indicates a partition of the graph into two subsets. The QAOA is a quantum algorithm for approximate optimization over bit strings. It has a depth parameter $p$ and performance can only improve as $p$ increases. The QAOA  produces a quantum state where the expectation of the MaxCut cost function is guaranteed to have a certain value.  This means that there must exist cuts of at least this value.  The MaxCut cost function is a sum of terms from each edge of the graph.  For  high-girth graphs, at each edge the neighborhood relevant to the QAOA  is a tree.  
See Fig.~\ref{fig:regular_tree} for an example at $p=3$.
To guarantee this we need $g\geq 2p+2$.  Each of these tree neighborhoods is isomorphic and at optimal parameters makes a contribution of $\tilde{c}_{\rm edge}(p)$ for each edge of the graph. The total contribution to the quantum expected value of the cost, at optimal parameters, is $|E(G)| \tilde{c}_{\rm edge}(p)$. The size of the largest cut must be at least the size of the cut found by the QAOA.  So we then have

\begin{equation}
M_g\geq \tilde{c}_{\rm edge}(p)
\end{equation}
for $g\geq 2p+2$. So $M_g$ is lower bounded by a quantum expectation value. 

Our paper explains how we calculate $\tilde{c}_{\rm edge}(p)$ with classical numerical computation.  We get values for $p$ up to $17$, corresponding to $g$ of $36$.
The numerical computations are highly accurate and the numbers we quote are good to at least four digits.   In fact we have  the best lower bound known to us for $M_g$ for any $g \geq 16$, improving on previous results by Thompson, Parekh, and Marwaha \cite{thompson2022explicit}.
At $p=17$ and $g=36$ we are guaranteed a cut fraction of $\tilde{c}_{\rm edge}(p\mathord{=}17) =\bestbound$. See Fig.~\ref{fig:mainResultTPM}.  Separately, from Refs.~\cite{csoka2015invariant} and \cite{gamarnik2018max}, one finds  $\lim_{g \to \infty} M_g \geq 0.912$.

Actually running the QAOA on a graph $G$ has time-complexity $O(|E(G)|)$, and it uses a constant-depth quantum circuit.
For any 3-regular graph of girth $g \geq 2p+2$, by using repeated applications of the QAOA, with total time-complexity $O(|E(G)|^2)$, we can obtain a cut of size at least $\floor{ |E(G)|\tilde{c}_{\rm edge}(p)}$ with probability $2/3$.
This may be viewed as an optimization or search problem with a promise (about the girth), with the goal of obtaining a cut of size above this threshold. There exist several classical algorithms for MaxCut, and we do not claim that the QAOA is superior in practice.  However, besides the QAOA, we do not know any quantum or classical algorithms proven to solve this promise problem in subexponential time.

 A natural point of comparison is the celebrated Goemans-Williamson (GW) algorithm \cite{goemans1995improved}.
The success of approximation algorithms is often measured by their approximation ratio: the ratio of the obtained value to the optimal value.
The GW algorithm guarantees an approximation ratio of 0.878 for general graphs, and an improved ratio of 0.932 when specialized to graphs of maximum degree 3~\cite{halperin2004max}.
Assuming the Unique Games Conjecture \cite{khot2002power}, it is NP-hard to exceed the GW algorithm's 0.878 guarantee for the worst case.
Our present analysis of the QAOA, however, focuses on cut fraction rather than approximation ratio.
While our highest cut fraction guarantee of $\bestbound$ immediately guarantees an approximation ratio at least this size, such an approximation ratio is already exceeded by the specialized GW algorithm. 
Meanwhile, the GW algorithm does not provide any direct guarantee on cut fraction.
So given a graph of girth $g \geq 36$, with the task of finding a cut of size at least $0.8971 |E(G)|$, the GW algorithm is not guaranteed to succeed in general, while our specification of the QAOA provides an efficient algorithm.  On the other hand, in the particular case of bipartite graphs (of any girth) the GW algorithm achieves the perfect cut, while the QAOA generally does not.

We briefly review QAOA, then describe the classical numerical methods used to obtain the performance guarantee.  
Our calculations are similar to the original calculations of Ref.~\cite{farhi2014quantum}, though performed at larger depth with distinct methods. 
We draw inspiration from the tensor network methods of Ref.~\cite{wurtz2021fixed}, where the same QAOA quantities were calculated, though at  lower depth $p$.   See also Refs.~\cite{basso_et_al:LIPIcs.TQC.2022.7, wybo2024missing} for related calculations at lower depth but including graphs of higher degree.  Our main technical contribution is to  analyze the QAOA for MaxCut on 3-regular graphs of large girth, using  modified numerical methods to allow higher depth than previously obtained. We then interpret these results in relation to previous lower bounds for the maximum cut and previous algorithms for approximating its value. Finally we apply the QAOA to the problem of Maximum Independent Set on high-girth 3-regular graphs. 

\section{Review of the QAOA}

The Quantum Approximate Optimization Algorithm (QAOA) is a quantum algorithm for finding approximate solutions to combinatorial optimization problems.  We seek to optimize an objective function $C$ over bit strings, which is usually a sum over terms that each involves a few bits.
The QAOA depends on an integer parameter $p \geq 1$ and produces quantum states of the form
\begin{equation}
|\paramv\rangle = U(B, \beta_p)U(C, \gamma_p)\cdots U(B, \beta_1)U(C, \gamma_1)|s\rangle.
\end{equation}
Here $|s\rangle$ is the uniform superposition over computational basis states, $U(C,\gamma) = e^{-i\gamma C}$ is diagonal in the computational basis, and $U(B,\beta) = e^{-i\beta B}$ where $B = \sum_j X_j$ is the sum of single-qubit Pauli X operators.
The angles $\gammav = (\gamma_1,\ldots,\gamma_p)$ and $\betav = (\beta_1,\ldots,\beta_p)$ are classical parameters  which specify the QAOA state.
For MaxCut, given a graph with vertices $V$ and edges $E$, the cost function operator counts edges crossing between two subsets in a partition,
\begin{equation}
\label{eq:cost_function}
C = \sum_{(j,k) \in E} \frac{1-Z_j Z_k}{2}.
\end{equation}
When measured in the computational basis, the state $|\paramv\rangle$ produces a bit string with a cost function value whose expectation is
\begin{equation}
F_p(\paramv) = \langle\paramv|C|\paramv\rangle.
\end{equation}
The angles $(\paramv)$ are optimized to maximize the expectation value $F_p(\paramv)$. The quantum circuit depth grows with $p$, and the performance at optimal parameters can only improve with  larger $p$.

\section{Pre-computing QAOA parameters for graphs with large girth}
 Traditionally, the QAOA is executed by applying the quantum circuit with initial parameters, estimating $F_p$ through measurements, then using classical optimization to update $(\gammav,\betav)$ and repeating the process. Here we take a different approach, emphasized especially by Ref.~\cite{wurtz2021fixed}, where parameters are chosen in advance for all graphs of a given girth. 

First note that the expectation value $F_p(\paramv)$ can be decomposed as a sum over edge terms. For MaxCut, each term depends only on the subgraph within distance $p$ of that edge. When a graph has girth at least 
\begin{align}
g \geq 2p+2
\end{align}
these neighborhoods are trees. Moreover, for all 3-regular graphs with this minimum girth, these trees are all isomorphic and make the same contribution to   $F_p(\paramv)$. We then can write

\begin{align}
F_p(\paramv)=|E(G)| ~\costperedge(\paramv)
\label{eq:FEc}
\end{align}
where 
\begin{align}
\label{eq:cost_per_edge}
    \costperedge(\paramv) =  \frac{1-\langle\paramv | Z_i Z_j |\paramv\rangle}{2}
\end{align}
and $(i, j) \in E(G)$ can be any pair of neighboring nodes in $G$ since they all make identical contributions.
Since \eqref{eq:FEc} holds for all 3-regular graphs of girth at least $g$, we have
\begin{align}
M_g\geq \costperedge(\paramv)
\end{align}
for any $\paramv$.  By finding optimal (or near optimal) parameters we improve the bound. We later discuss how we find the near optimal parameters $\approxparamv$.  Let
\begin{align}
 \tilde{c}_{\rm edge}(p) = c_{\rm edge}(\approxparamv).
 \end{align}
 Then we have
 \begin{align}
 \label{eq:Mc}
 M_g\geq \tilde{c}_{\rm edge}(p)
 \end{align}
 for any  $p \leq (g-2)/2$.

Wurtz and Lykov \cite{wurtz2021fixed} computed  $\tilde{c}_{\rm edge}(p)$ up to $p=11$ and we reproduce these numbers. We compute  up to $p= \highestp$.   More precisely, for any fixed $\paramv$, we compute  $\costperedge(\paramv)$ numerically using exact tensor network methods.
The numerical error in the computation of $\costperedge(\paramv) $ is negligible compared to the 4-digit accuracy reported in our results.
Then we maximize  to obtain some approximately optimal parameters $\approxparamv$.  Despite this approximate optimization, the value  $\tilde{c}_{\rm edge}(p)$ then serves as a precise lower bound for $M_g$, due to \eqref{eq:Mc}.

When running the QAOA for graphs of girth greater than  36, it would be helpful to pre-compute the optimal parameters for $p > \highestp$, which we have not done.  For sufficiently large $p$ this may be intractable, or it may be well-approximated by extrapolation of the optimal parameters from smaller $p$ to larger $p$.  Regardless, for the purpose of proposing a formal algorithm, for all 3-regular graphs of girth greater than  36, we propose running the $p=\highestp$ QAOA with our fixed, pre-computed parameters.  This protocol already gives an algorithm with polynomial runtime with respect to graph size, with the best currently known performance guarantee on cut fraction.  The only other algorithms we currently know with the same performance guarantee take exponential time.  One example is the brute force algorithm, which finds the best cut in exponential time, and is therefore guaranteed to find a cut of size at least $0.8971 |E(G)|$, whose existence follows from our present results. A classical algorithm of Williams~\cite{williams2005new} finds the best cut in exponential time but  using a smaller exponent than brute force search.

\section{Analyzing time-complexity}

Here we detail the runtime of the QAOA as an algorithm running on a quantum computer to find an actual cut. (This analysis is separate from the classical numerical computations discussed in 
Section \ref{sec:methods}, which establish performance guarantees and serve as a pre-computation that is independent of problem instance.)
 We focus on the QAOA for $d$-regular graphs on $|V(G)|$ vertices, treating $d$ as a constant.  The $MaxCut$ problem size is the number of edges, $|E(G)|=(d/2) |V(G)|$.
The QAOA at depth $p$ uses $O(p |E(G)|)$ total gates. When analyzing the runtime of quantum algorithms, we are interested in both the total number of gates and also the circuit depth, which counts the minimal number of layers such that each layer involves non-overlapping gates.

\begin{figure}
    \centering
    \includegraphics[width=0.7\linewidth]{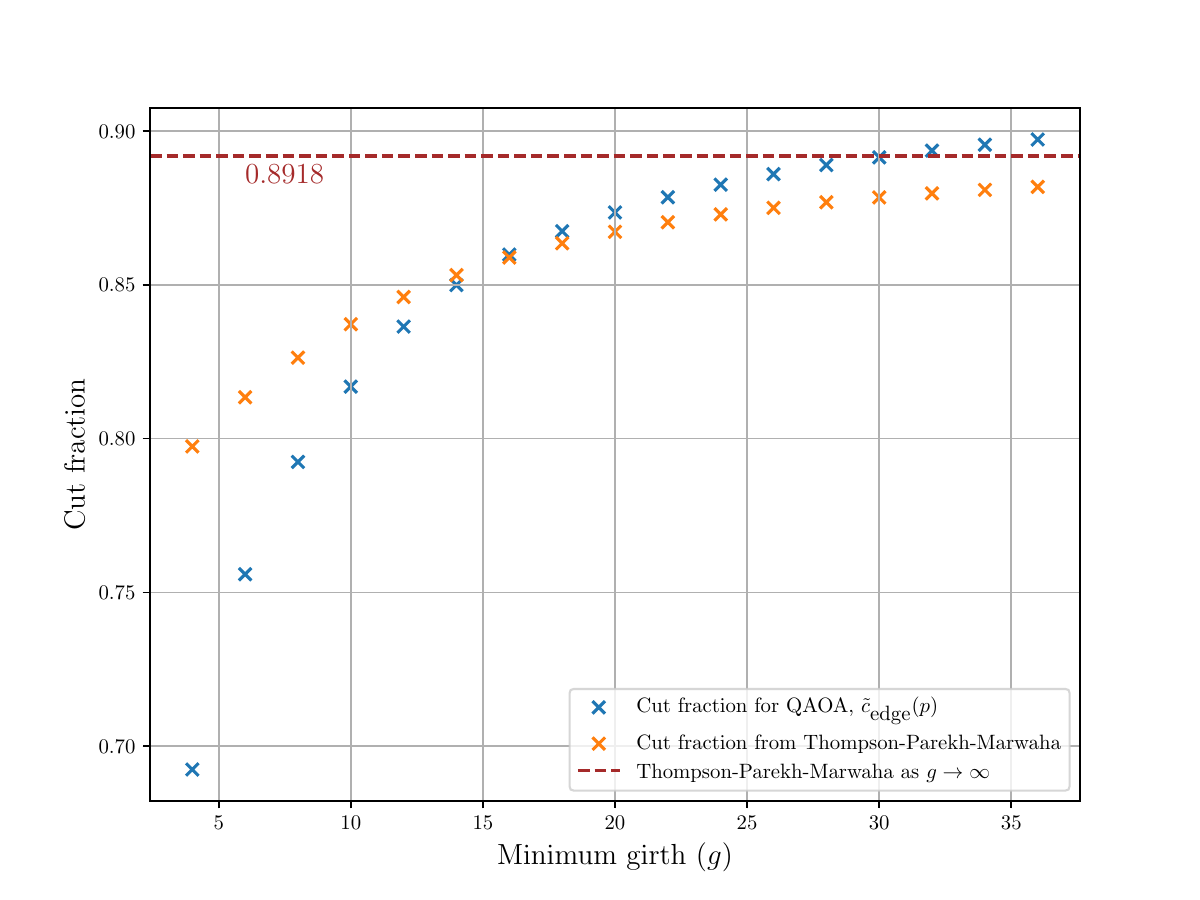}
    \caption{Lower bound on  $M_g$ for 3-regular graphs given by the present work (blue crosses). As a comparison we show the results of Thompson-Parekh-Marwaha (TPM)
\cite{thompson2022explicit} in orange. The QAOA lower bound exceeds the TPM lower bound at $g\geq 16$ (corresponding to $p \geq 7$).
    The QAOA lower bound exceeds the TPM $g \to \infty$ bound at $g\geq 32$ ($p \geq 15)$.
 }
    \label{fig:mainResultTPM}
\end{figure}

\begin{figure}
    \centering
    \includegraphics[width=1.0\linewidth]{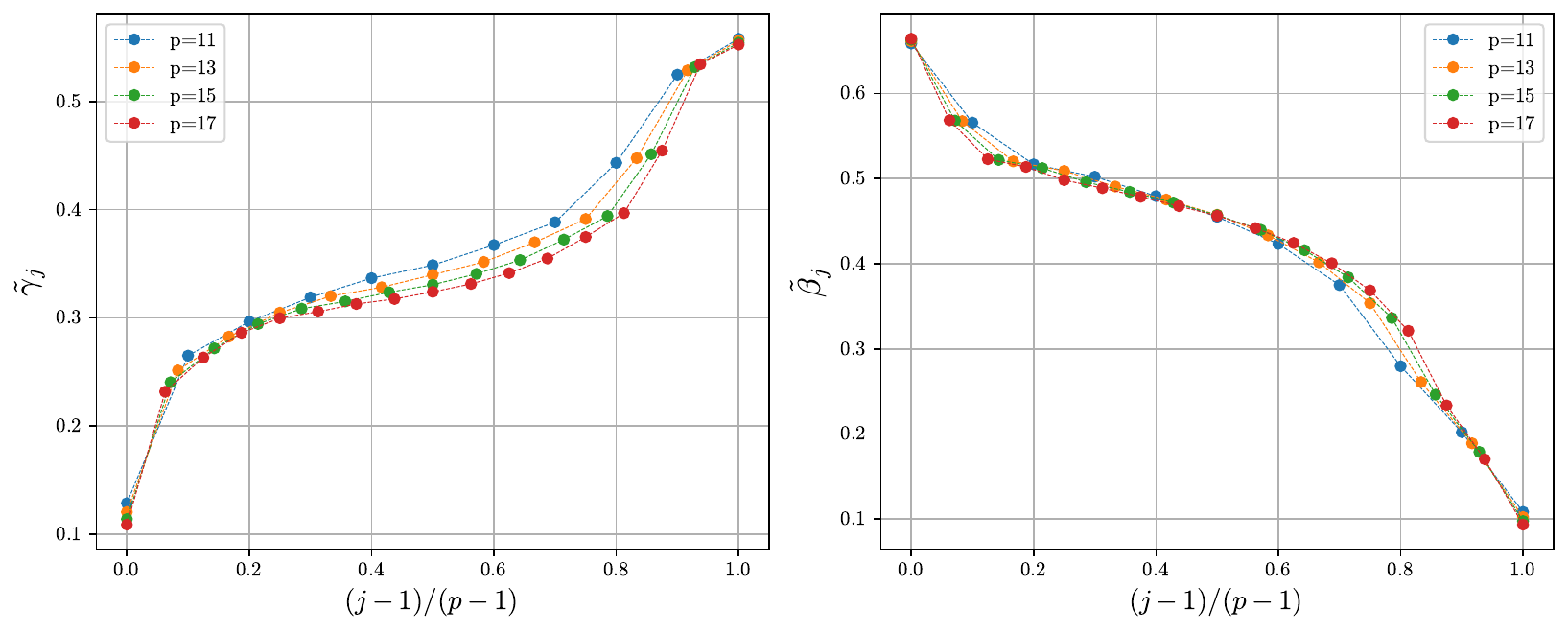}
    \caption{Optimized parameters $\approxgammav$ and $\approxbetav$ as a function of $(j-1)/(p-1)$ for different values of $p$ up to p=17.
    These figures suggest that the optimal parameters might approach fixed curves as $p$ increases.}
    \label{fig:parameters}
\end{figure}

While the parameter $p$ is sometimes called the ``depth'' of the QAOA, its relation to circuit depth depends on the graph.  The unitary $U(B,\beta)$  can be implemented in a single layer of depth 1. 
On the other hand, while $U(C,\gamma)=e^{-i \gamma C}$ is naturally implemented by a two-qubit gate for every edge, these gates overlap whenever two edges share a vertex, so $U(C,\gamma)$ requires circuit depth larger than 1. Szegedy noted that one can use Vizing's theorem to calculate the circuit depth \cite{szegedy2019qaoa}.  For graph $G$, the ``edge chromatic number'' $\chi'(G)$ is the minimal number of colors required for an edge-coloring where no two adjacent edges share the same color.  By using a layer of gates for each color, we see $U(C,\gamma)$ can be implemented in depth $\chi'(G)$ using 2-local gates.
Vizing's theorem states that for general graphs, $\chi'(G) \leq \Delta(G) + 1$, where $\Delta(G)$ is the maximum degree. Again treating $d$ as a constant, we conclude $U(C,\gamma)$ has $O(1)$ circuit depth, and the full QAOA has circuit depth $O(p)$.  
To achieve this $O(p)$-depth circuit, we need to classically pre-compute to actually find one of the colorings guaranteed by Vizing's theorem.  For a bounded degree graph, this computation runs in time $O(|E(G)|)$ \cite{bernshteyn2023fast}, the same as the problem size, hence not contributing to the asymptotic complexity.

In this section,  when we analyze the time complexity of running the QAOA on a quantum computer  at fixed $p$,  we imagine running with fixed parameters $\paramv$ given in advance. There is no outer loop variational search.    The quantum circuit  uses $|V(G)|$ qubits, with $O(|E(G)|)=O(|V(G)|)$ gates and constant circuit depth.

After running the QAOA circuit and performing a measurement in the computational basis, we obtain a cut of the graph that depends on the measurement outcome.  The expected value of the cut size is $|E(G)| \tilde{c}_{\rm edge}(p)$. We can also ask for a guarantee that the cut is above (the floor of) this expected value with high probability.  To that end, repeat the QAOA circuit and measurement many times and record the best outcome.  The probability of sampling a cut size of at least $\floor {|E(G)| \tilde{c}_{\rm edge}(p)}$ with a single sample is at least $1/|E(G)|$. 
To see this, let $C\in\{0,1,\ldots,|E(G)|\}$ denote the cut size obtained from a single sample, and denote $C_0=\floor{ |E(G)|\,\tilde c_{\rm{edge}}(p)}$, with $p_0 := \Pr\!\big(C \ge C_0\big)$.  We calculate
\begin{equation}
\begin{split}
\mathbb{E}[C]
& \le (C_0-1)\Pr(C \leq C_0-1) + |E(G)|\,\Pr(C\ge C_0) \\
& = (1-p_0)(C_0-1) + p_0 \,|E(G)| \\
& \leq C_0 - 1 + p_0 |E(G)|,
\end{split}
\end{equation}
where the first line uses the fact $C$ takes integer values with $C \leq |E(G)|$. Combining the above with $\mathbb{E}[C] \geq  C_0$, we obtain $p_0 \geq 1/|E(G)|$.
Then with $O(|E(G)|)$ repetitions of the QAOA circuit and measurement, we obtain a sample with cut at least $\floor {|E(G)| \tilde{c}_{\rm edge}(p)}$, with probability at least $2/3$ (or any constant probability below 1).

\section{Numerical results and methods}
\label{sec:numerics}

\subsection{Results}
\label{sec:results}

\begin{table}[ht]
\centering
\begin{tabular}{cccccccccc}
\hline
\multicolumn{1}{|c|}{$p$} & 1 & 2 & 3 & 4 & 5 & 6 & 7 & 8 & \multicolumn{1}{c|}{9} \\ \hline
\multicolumn{1}{|c|}{$\tilde{c}_{\rm edge}$} & $0.6924$ & $0.7559$ & $0.7923$ & $0.8168$ & $0.8363$ & $0.8498$ & $0.8597$ & $0.8673$ & \multicolumn{1}{c|}{$0.8734$} \\ \hline 
\\
\cline{1-9}
\multicolumn{1}{|c|}{$p$} & 10 & 11 & 12 & 13 & 14 & 15 & 16 & \multicolumn{1}{c|}{17} \\ \cline{1-9}
\multicolumn{1}{|c|}{$\tilde{c}_{\rm edge}$} & $0.8784$ & $0.8825$ & $0.8859$ & $0.8888$ & $0.8913$ & $0.8935$ & $0.8954$ & \multicolumn{1}{c|}{$0.8971$} \\ \cline{1-9}
\end{tabular}
\caption{$\costperedge$ at optimized angles $\approxgammav$ and $\approxbetav$ as a function of $p$ up to $p=17$.}
\label{table:optimal_c_edge}
\end{table}

In order to maximize the quantum expected value of the cut size, we numerically evaluate the quantum expectation value of the cost function of Eq.~\eqref{eq:cost_function}.  As we emphasized before, at the QAOA depths we consider, on these high girth graphs, each edge has an isomorphic tree neighborhood and so each edge gives the same contribution.  
We find  values of $(\paramv)$ that approximately maximize $\costperedge(\paramv)$.  For fixed $(\paramv)$ the quantum expectation is evaluated using 
the tensor network contraction method described in Section~\ref{sec:methods}.
The blue crosses in Fig.~\ref{fig:mainResultTPM}  show the optimized values of $\costperedge(\paramv)$ at approximately optimal parameters, $\tilde{c}_{\rm edge}(p) = c_{\rm edge}(\approxparamv)$. The values are listed in
Table~\ref{table:optimal_c_edge}.
The parameters $\approxgammav$ and $\approxbetav$ are shown in Fig.~\ref{fig:parameters} for different values of $p$.

\subsection{Methods}
\label{sec:methods}

\begin{figure}[t]
    \centering
    (a)
    \includegraphics[width=0.45\linewidth]{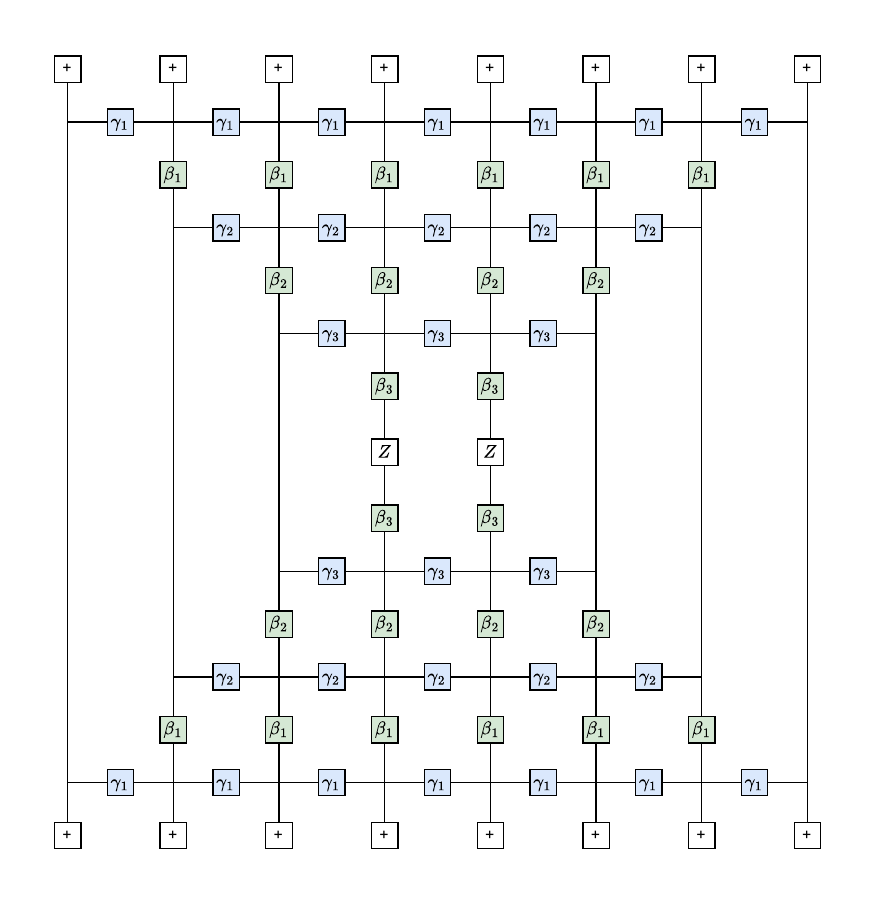}
    (b)
    \includegraphics[width=0.45\linewidth]{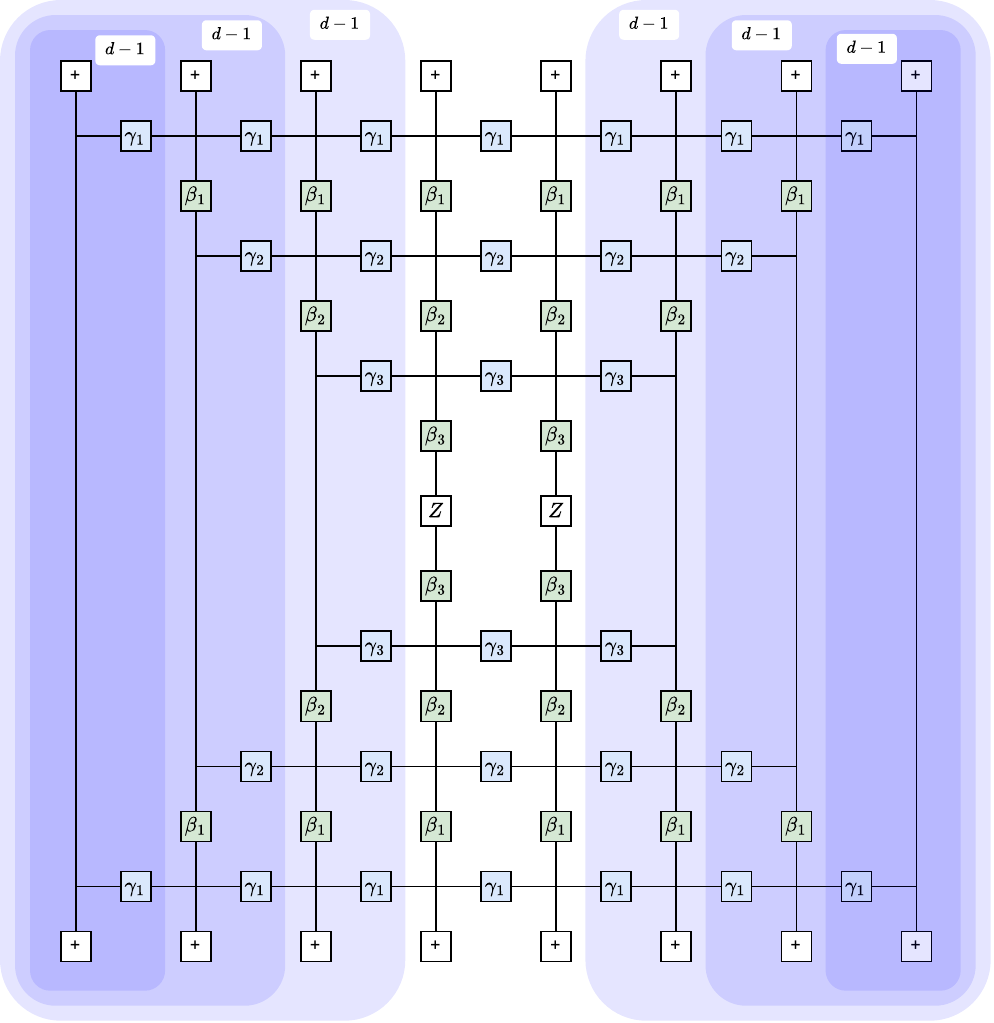}
    \caption{Tensor network for the computation of $\langle \paramv| Z_i Z_j |\paramv \rangle$ for $p=3$. On the left (a) we have   $d=2$ and on the right (b) arbitrary $d$.
    In the latter case, the tensor network diagramatic notation is extended to denote extry-wise exponentiation of a tensor.
    In particular, the result of contracting the tensors in one of the colored boxes is raised to the power $(d-1)$ before proceeding to later contractions, as expressed in Eq.~\eqref{eq:tensor_power}.
    The time and space complexities of the contraction performed in this way are both $\mathcal{O}(2^{2p})$.
    The complexity does not depend on $d$.
    Higher values of $p$ are tackled in a similar fashion.}
    \label{fig:TN-tree}
\end{figure}

In order to evaluate $\costperedge(\paramv)$, we consider the operator $Z_i Z_j$, as well as all quantum gates in the QAOA circuit contained in the light cone of this operator.
In the simple case of $d=2$, i.e., when $G$ is a line, the tensor network that evaluates $\langle Z_i Z_j \rangle$ is that of Fig.~\ref{fig:TN-tree}(a).
Here we have followed the tensor definitions:
\begin{align}
    \label{eq:tensor_definitions}
    \vcenter{\hbox{\includegraphics[width=0.08\linewidth]{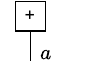}}}
    &= \quad
    \frac{1}{\sqrt{2}} \quad \forall a
    &\vcenter{\hbox{\includegraphics[width=0.16\linewidth]{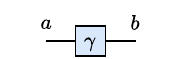}}}
    &= \quad
    \begin{cases}
    e^{i \gamma} &{\rm if} \,\, a=b \\
    e^{-i \gamma} &{\rm if} \,\, a\neq b
    \end{cases}
    \nonumber \\
    \vcenter{\hbox{\includegraphics[width=0.08\linewidth]{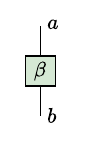}}}
    &= \quad
    \begin{cases}
    \cos{(\beta)} &{\rm if} \,\, a=b \\
    -i \sin{(\beta)} &{\rm if} \,\, a\neq b
    \end{cases}
    &\vcenter{\hbox{\includegraphics[width=0.08\linewidth]{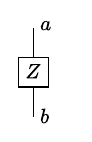}}}
    &= \quad
    \begin{cases}
    1 &{\rm if} \,\, a=b=0 \\
    -1 &{\rm if} \,\, a=b=1 \\
    0 &{\rm otherwise}
    \end{cases}
\end{align}
where all tensor indices have support $\{0, 1\}$.
Note that every gate of the form $e^{i \gamma Z_i Z_j}$ is diagonal, which we explicitly exploit in writing its corresponding tensor over only two variables and making use of hyperindices in the tensor network, corresponding to hyperedges in the underlying network.

\begin{figure}[t]
    \centering
    \includegraphics[width=0.8\linewidth]{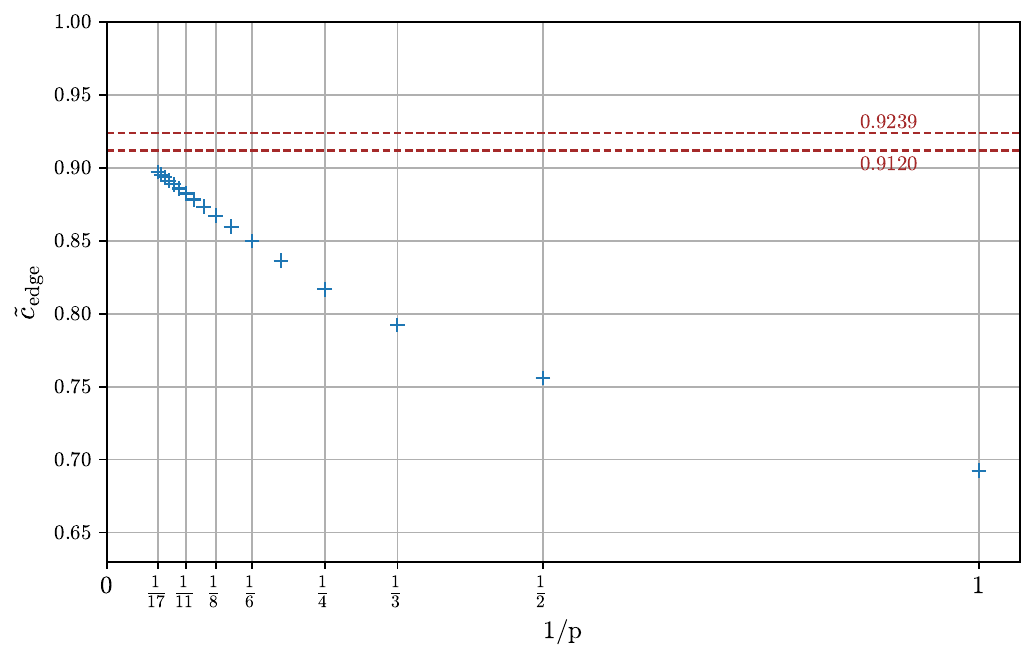}
    \caption{
    We plot $ \tilde{c}_{\rm edge}$ which is $\costperedge$ at optimized angles $\approxgammav$ and $\approxbetav$ as a function of $1/p$.  
    The value 0.912 is the lower bound on $\lim_{g \to \infty} M_g$ given by Refs. \cite{csoka2015invariant,gamarnik2018max}.
    If  $\tilde{c}_{\rm edge}$ were to exceed this value at larger $p$, it would provide a new lower bound on $\lim_{g \to \infty} M_g$.
    The value 0.9239 is the upper bound on the expected cut fraction of large random $3$-regular graphs given in \cite{harangi2025rsb}, and $\tilde{c}_{\rm edge}$ cannot exceed this value.
    }
    \label{fig:QAOACutFraction}
\end{figure}

In the generic case of $d>2$ the number of qubits in the light-cone of the $Z_i Z_j$ operator grows exponentially with $p$.
For a 3-regular graph (\emph{i.e.} $d=3$), at $p=1, 2, 3, \ldots$ the size of the light cone is $6, 14, 30, \ldots$
In general, for a $d$-regular graph, the number of qubits at depth $p$ is
$ 2 \frac{(d-1)^{p+1}-1}{d-2}$.
For $d=3$ and $p=17$ the size of the light-cone is 524,286 qubits.
However, due to the fact that all branches in the regular tree are identical, we can perform this computation in a compact way that is not affected by the exponential growth in the number of qubits in the calculation.
We contract a single branch inwards towards the root of the tree, raising its tensor entries to the $(d-1)$th power before proceeding to the next level of the tree.
This is expressed in graphical notation in Fig.~\ref{fig:TN-tree}(b), where we have made use of the definition
\begin{align}
    \label{eq:tensor_power}
    \vcenter{\hbox{\includegraphics[width=0.18\linewidth]{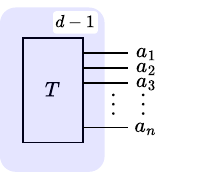}}}
    \equiv \quad
    \left(T_{a_1, a_2, \ldots, a_n}\right)^{d-1}
\end{align}
for the entrywise exponentiation of a tensor.
Note that the cost of the evaluation of Eq.~\eqref{eq:cost_per_edge} as expressed in Fig.~\ref{fig:TN-tree}(b) has both time and space complexities that grow as $\mathcal{O}(2^{2p})$.
This is quadratically better than the time complexity reported in Refs.~\cite{basso_et_al:LIPIcs.TQC.2022.7}~and~\cite{wybo2024missing} for finite $d$.
Note also that the cost is independent of $d$, in contrast to that of the method used in Ref.~\cite{wurtz2021fixed}, which scales exponentially in $d$ and was run  up to $p=11$ at $d=3$ and lower values of $p$ at larger $d$.
This allows us to evaluate and optimize Eq.~\eqref{eq:cost_per_edge} up to $p=17$ for any value of $d$.
Our implementation of the method is written in C++ and parallelized using OpenMP~\cite{dagum1998openmp}.
We also make use of the Eigen library for the manipulation of vectors~\cite{eigenweb} as well as the LBFGS++ library, which implements the Limited-memory BFGS algorithm for unconstrained optimization problems~\cite{lbfgspp}.

We remind the reader that all of our numerical methods are exact.  We use computers to evaluate expressions, but there are no approximations.
For $d=3$ and at depth $p$ we evaluate $\costperedge(\paramv)$ on a tree with $(2^{(p+2)} - 2)$ vertices at a computational cost of $O(4^p)$.
We go to p of 17.
An alternate approach would be to use a quantum computer with $(2^{(p+2)} -2)$ qubits.
Here we would evaluate the central edge of the tree using the full QAOA cost.
Although we would be running a quantum computer, we would not be finding cuts but rather estimates of lower bounds on MaxCut values. It appears that this scales more favorably than our exact calculation of lower bounds, at least  while $d \leq 4$. For this method to go beyond our $p=17$ result  requires a highly accurate quantum computer capable of handling  graphs with at least one million vertices. Repeated measurements would extract an accurate estimate of the quantum expectation but not a proof of a lower bound.

\section{MaxCut Comparisons}

We can directly compare our results with those of Thompson,  Parekh and Marwaha 
\cite{thompson2022explicit}. They have a classical algorithm for MaxCut which they apply to large girth graphs.  They lower bound the performance of their algorithm to obtain lower bounds on $M_g$ for all $g$. Using the formula found in Theorem 1 of their paper we can compare with our results. See Fig.~\ref{fig:mainResultTPM}.   For any $g \geq 16$ the QAOA (at appropriate $p$) finds cuts bigger than the TPM guarantees. Note that the limit as $g$ goes to infinity of their bound is $0.8918$ which we exceed at $g$ of $32$.

The results of Cs{\'o}ka et al \cite{csoka2015invariant} and Gamarnik and Li \cite{gamarnik2018max} imply that the limit as $g$ goes to infinity of $M_g$ is greater than or equal to 0.912. (These methods could in principle be  used to find bounds on $M_g$ for finite $g$, but as far as we know this has not been done.)  Our numerical techniques are stretched to the limit at $p$ of $17$ and we do not have an analytic way of taking $p$ to infinity. See Fig.~\ref{fig:QAOACutFraction}  where we plot $\tilde{c}_{\rm edge}$ versus $1/p$ and also show the target of 0.912.  The reader can decide if we pass this target as $p$ increases, an empirical question left for future computations.

The graphs which we have considered have large girth and every edge sits in a tree neighborhood.  We can ask about large random 3-regular graphs where almost all edges sit in tree neighborhoods.  Such graphs have an order unity number of triangles, squares, pentagons, or any cycle of constant length,  which means that they are not strictly large girth.  However, since there are so few short cycles, any lower bound on $M_g$ for any $g$ is a lower bound on the cut fraction of a typical large random 3-regular graph \cite{gamarnik2018max}. Furthermore the QAOA at level $p$ achieves cut fraction $\tilde{c}_{\rm edge}(p)$ on these graphs.
There is an upper bound on  the cut fraction of typical large random 3-regular graphs of 0.9239~\cite{harangi2025rsb}, building on Refs.~\cite{coja2022ising, zdeborova2010conjecture, kardovs2012maximum,mckay1982maximum}.
Therefore the QAOA cut fraction $\tilde{c}_{\rm edge}(p)$ cannot exceed 0.9239 even as $p \to \infty$.  This value is shown in Fig.~\ref{fig:QAOACutFraction}. 

\section{Maximum Independent Set}

An independent set is a subset of the vertices of a graph with the property that there are no edges between any two members of the subset.  The Maximum Independent Set problem (MIS) is to find the biggest independent set. The MIS problem is NP-hard for general graphs.  Here we will use a general connection between independent sets and MaxCut that allows us to use our MaxCut results to get quick bounds on the independence ratio which is the fraction of vertices in the largest independent set.  We then apply a variant of the QAOA to large-girth 3-regular graphs to establish better bounds on the independence ratio. We are not aware of any other bounds for MIS on this class of  fixed girth  graphs.  However there are results in the limit as the girth goes to infinity~\cite{csoka2016independent}.

\begin{table}[b]
\centering
\begin{tabular}{cccccccccc}
\hline
\multicolumn{1}{|c|}{$p$} & 1 & 2 & 3 & 4 & 5 & 6 & 7 & 8 & \multicolumn{1}{c|}{9} \\ \hline
\multicolumn{1}{|c|}{Indep. ratio (2 params.)} & $0.2693$ & $0.3169$ & $0.3442$ & $0.3626$ & $0.3772$ & $0.3874$ & $0.3948$ & $0.4005$ & \multicolumn{1}{c|}{0.4051} \\ \hline 
\multicolumn{1}{|c|}{Indep. ratio (3 params.)} & $0.2852$ & $0.3324$ & $0.3591$ & $0.3749$ & $0.3861$ & $0.3942$ & $0.4005$ & $0.4054$ & \multicolumn{1}{c|}{0.4094} \\ \hline 
\\
\cline{1-9}
\multicolumn{1}{|c|}{$p$} & 10 & 11 & 12 & 13 & 14 & 15 & 16 & \multicolumn{1}{c|}{17} \\ \cline{1-9}
\multicolumn{1}{|c|}{Indep. ratio (2 params.)} & $0.4088$ & $0.4118$ & $0.4144$ & $0.4166$ & $0.4185$ & $0.4201$ & $0.4216$ & \multicolumn{1}{c|}{0.4228} \\ \cline{1-9}
\multicolumn{1}{|c|}{Indep. ratio (3 params.)} & $0.4126$ & $0.4154$ & $0.4177$ & $0.4197$ & $0.4215$  & $0.4230$ & \multicolumn{1}{c|}{0.4244} \\ \cline{1-8}
\end{tabular}
\caption{Independence ratio achieved by the QAOA for various values of $p$.}
\label{table:optimal_independence_ratio}
\end{table}

A convenient local cost function for the MIS problem is 
\begin{equation}
\label{eq:I_1}
I_1 = 
\sum_{i \in V} b_i - \sum_{\langle i,j\rangle \in E} b_i b_j,
\end{equation}
written as a function of the bit string $(b_1, \ldots,b_{|V|})$, where each bit string also represents a subset of the vertex set $V$, with $b_i=1$ if vertex $i$ is in the subset. Note this cost function is defined on all subsets and not just independent sets. The first term of $I_1$ is the Hamming weight which we want to make big, and the subtracted term counts violations of the independent set property. 
We now argue that given any bit string with associated cost $I_1$, 
there exists an independent set of size at least $I_1$. In particular, we consider the set of vertices labeled 1 in the bit string, and we construct a subset of these vertices that forms an independent set of size at least $I_1$.
To begin, given a bit string with cost $I_1$, assume that the set of $1$'s is not yet an independent set. Choose any violated edge $\langle i, j \rangle$, that is with $b_i=b_j=1$, then  choose either $i$ or $j$  and re-label it as $0$, that is, remove the vertex from the set under consideration.  Consider the cost function $I_1$ for this new string: the bit flip decreases the Hamming weight by $1$ and decreases the number of violations by at least $1$, so $I_1$ cannot decrease.  Repeat this until there are no more violations so the final bit string must have $1$'s that form a valid independent set.  The new value of $I_1$ is equal the size of the independent set which must be at least the original value of $I_1$. Now the maximum of $I_1$ occurs when $b$ corresponds to an actual maximum independent set so achieving a high value of $I_1$ gives an approximation to the size of the largest independent set.

We now  use the connection between MaxCut and independent sets to get a general bound on MIS. Consider the problem of maximizing the total number of vertices in two disjoint independent sets. We consider a  cost function given by the sum of two terms, one for each set:

\begin{equation}
\label{eq:I_2}
I_2= \left[ \sum_{i \in V} b_i - \sum_{\langle i,j\rangle \in E} b_i b_j \right] + \left[ \sum_{i \in V} (1-b_i) - \sum_{\langle i,j\rangle \in E} (1-b_i)(1- b_j ) \right].
\end{equation}
Write the two terms as $I_2=I_{2,1}+I_{2,0}$. For any input bit string, the $1$'s represent one set and the $0$'s the complementary set. (Note that $0$ and $1$ here do not mean what they meant in the previous paragraph.)  By the above argument regarding $I_1$, first applied to $I_{2,1}$, there exists a subset of $1$'s that forms an independent set, with size at least $I_{2,1}$.  Meanwhile, the same argument applied to $I_{2,0}$ implies there exists a subset of $0$'s that forms an independent set, with size at least $I_{2,0}$. Thus we have two disjoint independent sets of total size at least $I_2$. 

 We can rewrite Eq.~\eqref{eq:I_2} to get
 \begin{equation}
 I_2=|V|-|E|+\sum_{\langle i,j\rangle \in E}(b_i+b_j-2b_ib_j)
 \end{equation}
 where the sum on edges is the MaxCut cost. Let $w$ be the maximum fraction of combined vertices in two disjoint independent sets.  Then for any graph,
\begin{equation}
w\geq \frac{I_2}{|V|} = 1-\frac{|E|}{|V|} + \frac{C(b)}{|V|}
\end{equation}
where $C(b)$ is the cut size associated to any bit string $b$.  For 3-regular graphs we get
\begin{equation} \label{eq:w1}
w\geq \frac{3}{2}\frac{C(b)}{|E|}-\frac{1}{2}.
\end{equation}

As an aside, we note an interesting relation between $w$ and $\mu$, where $\mu$ denotes the cut fraction of the best cut:
\begin{equation} \label{eq:mu-w}
w = \frac{3}{2}\mu-\frac{1}{2}.
\end{equation}
To see this, first note
\begin{equation}
w\geq\frac{3}{2}\mu-\frac{1}{2},
\end{equation}
which follows immediately from Eq.~\eqref{eq:w1}. This holds for any 3-regular graph with no restriction on girth.  Now we refer to  Gamarnik and Li \cite{gamarnik2018max}. From the argument in their Section 6, one can deduce
\begin{equation}
\mu \geq \frac{2}{3}w+\frac{1}{3}.
\end{equation}
Together, these two inequalities yield the claimed Eq.~\eqref{eq:mu-w} and we end our aside.

\begin{figure}[t]
    \centering
    \includegraphics[width=0.8\linewidth]{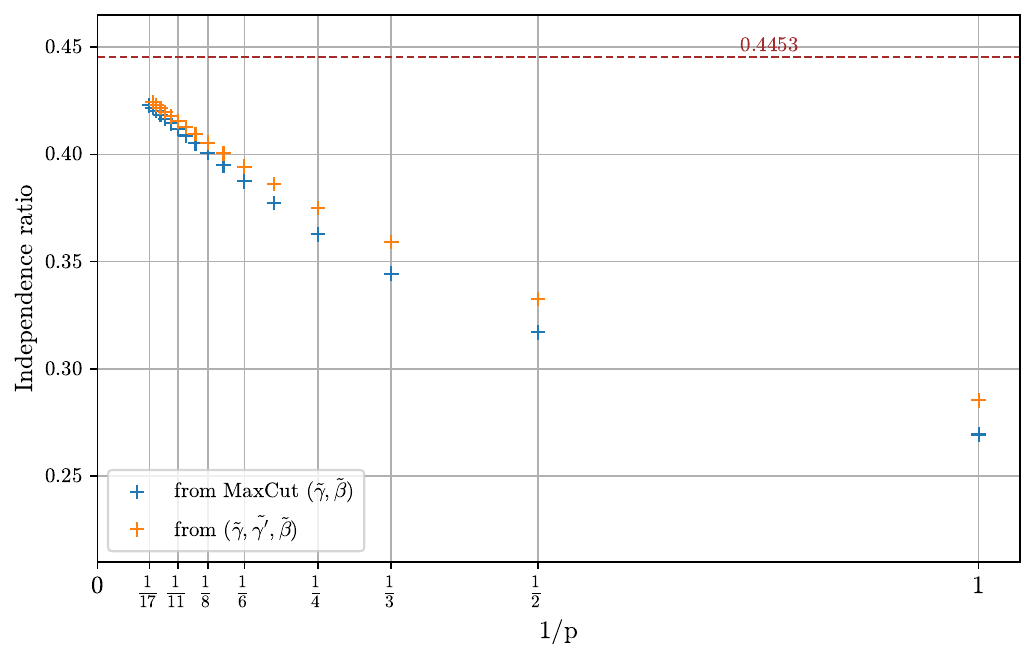}
    \caption{
   We plot the independence ratio for the MIS problem as a function of $1/p$.  
   The value 0.4453 is the best available bound on the independence ratio for large-girth 3-regular graphs Ref.~\cite{csoka2016independent}.
   We show bounds on the independence ratio for angles $\gammav$ and $\betav$ optimized for MaxCut, as well as the set of parameters $(\gammav, \boldsymbol{\gamma^\prime}, \betav)$ (see expression~\eqref{eq:two_angle_driver}).
   In the latter case we necessarily achieve  larger values of the independence ratio, although the improvement over the $(\paramv)$ alternative decreases with $p$.
   }
    \label{fig:QAOAIndependenceRatio}
\end{figure}

Let $ir$ be the independence ratio, the fraction of vertices in the largest independent set.  Then $ir \geq w/2$, since one of the two independent sets must contain at least half of their combined vertices.  So from Eq.~\eqref{eq:w1} we have
\begin{equation}
\label{eq:MIS7}
ir\geq \frac{3}{4}\frac{C(b)}{|E|}-\frac{1}{4}.
\end{equation}
Taking the quantum expectation and using our MaxCut result for large girth 3-regular graphs we have
\begin{equation}
\label{eq:MIS9}
   ir \geq  \frac34 \tilde{c}_{\rm edge}(p) - \frac14  \,. 
\end{equation}
So we now have a lower bound on the independence ratio $ir$ for any $g\geq 2p+2$ using the values from Table~\ref{table:optimal_c_edge}.   These values are given as the first line in Table~\ref{table:optimal_independence_ratio}. In particular for girth at least $36$ we get a lower bound of $0.4228$.

But we can do better. Return  to the cost function of Eq.~\eqref{eq:I_1}.   First we need to write it as a sum over edges. For 3-regular graphs,
\begin{equation}
    I_1 = \sum_{\langle i,j\rangle \in E} \left[ \frac13 (b_i + b_j) - b_i b_j\right].
\end{equation}
Now using the fact that $b_i=(1-z_i)/2$ we can write the cost in terms of the $Z_i$ operators which is more convenient,
\begin{equation}
\label{eq:MIS0}
    \sum_{\langle i,j\rangle \in E} \left[\frac12 \left(\frac12-\frac12 Z_i Z_j\right) - \frac16 + \frac{1}{12}(Z_i +Z_j) \right].
\end{equation}
 Again we are looking at graphs with $g\geq 2p+2$ so each edge makes an identical contribution to the quantum expectation in the QAOA state. The quantum expectation equals the contribution from any edge times the number of edges $|E|$, that is,
\begin{equation}
\label{eq:MIS1}
    \langle \psi|\left[\frac12 \left(\frac12-\frac12 Z_i Z_j\right) - \frac16 + \frac{1}{12}(Z_i + Z_j) \right] | \psi \rangle \times |E|
\end{equation}
 where as before $\langle i,j\rangle$ is any edge in the graph, and $|\psi\rangle$ is some state produced by the QAOA.

 We are free to use any local operator to drive the QAOA. Here we modify the cost function part of the driver and but not the sum of the $X$'s.  We will introduce two parameters, $\gamma$ and $\gamma'$, for each layer of the cost function unitary.  In particular for the driving cost function operator we try

\begin{equation}
\label{eq:two_angle_driver}
\gamma \sum_{\langle i,j\rangle \in E} Z_i Z_j+\gamma'\sum_{i \in V} Z_i.
\end{equation}
Including the $\beta$ at each layer there are a total of $3p$ parameters. If $\gamma'=0$, this amounts to using the MaxCut cost function as the driver.
Using symmetry we can show that the linear term in the objective in expression~\eqref{eq:MIS1} vanishes. So in fact with $\gamma'=0$ both the driver and objective function are those of MaxCut up to constants. Noting that $|E|=\frac{3}{2}|V|$  and dividing ~\eqref{eq:MIS1} by $|V|$ gives, at optimal parameters, the right hand side of ~\eqref{eq:MIS9}. If the ratio of $\gamma$ to $\gamma'$ is set to the right constant, see expression~\eqref{eq:MIS0}, then the driver is the same as the objective function. So if we optimize over all $\gamma$'s and $\gamma'$'s and $\beta$'s we do at least as well as using the MaxCut cost function as the driver or the local MIS cost function as the driver.  In Table~\ref{table:optimal_independence_ratio} we give lower bounds on the independence ratio for MIS produced by this optimization over $3p$ parameters.

\begin{figure}
    \centering
    \includegraphics[width=1.0\linewidth]{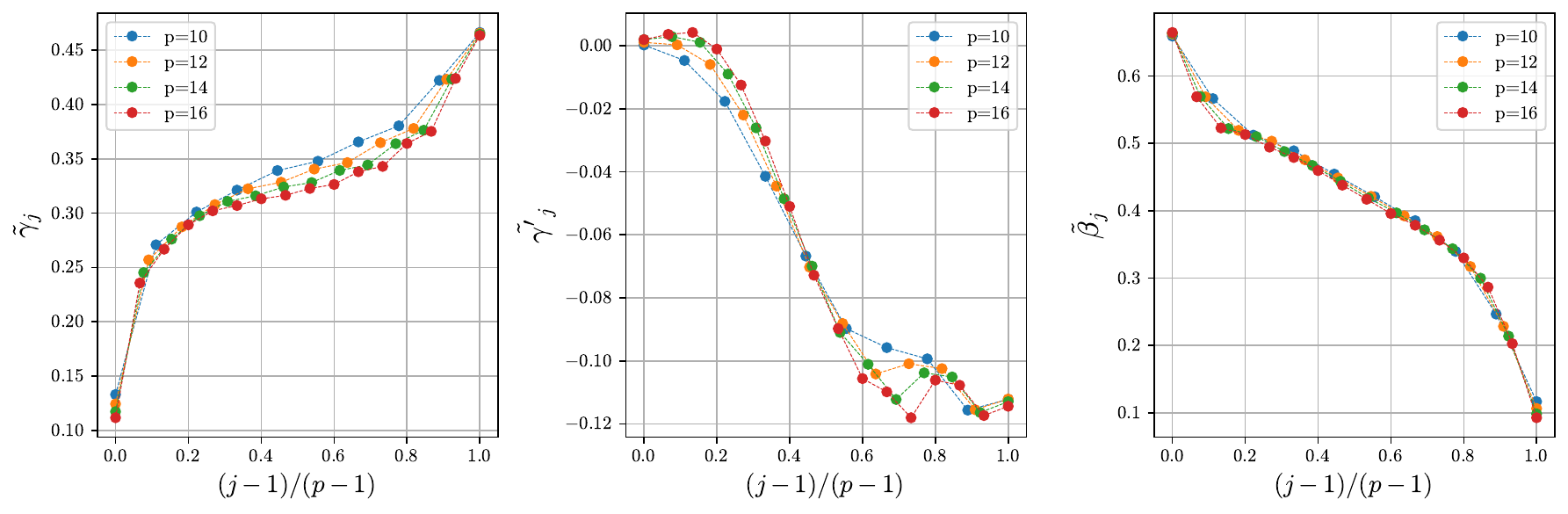}
    \caption{Optimized parameters $\approxgammav$, $\boldsymbol{\tilde{\gamma^\prime}}$ and $\approxbetav$ as a function of $(j-1)/(p-1)$ for different values of $p$ up to p=15 for the MIS problem.}
    \label{fig:parameters_mis}
\end{figure}

In Fig.~\ref{fig:QAOAIndependenceRatio} we show the lower bounds we get on the independence ratio as a function of $1/p$ from our MaxCut results and using the $3$-parameter per layer ansatz. The $3$-parameter ansatz must lie above the other but the difference shrinks as $p$ grows. The red line is the best available bound on the independence ratio for large girth 3-regular graphs as the girth goes to infinity \cite{csoka2016independent}.
The optimized parameters $(\tilde{\gammav}, \tilde{\boldsymbol{\gamma^\prime}}, \tilde{\betav})$ are shown in Fig.~\ref{fig:parameters_mis} for different values of $p$.

To find independent sets which meet these guarantees you can run the QAOA on quantum hardware with constant circuit depth for a constant $p$. This gives a polynomial-time quantum algorithm for this problem. We know of no classical algorithm that performs as well (in terms of guaranteed independence ratio) on the problem of MIS on 3-regular graphs of girth at least $g$.  

\section{Conclusions}

We  proved a new lower bound for MaxCut on high-girth graphs by using a classical computer to analyze the performance of a quantum algorithm.  This graph-theoretic result holds even if quantum computing is infeasible. We are unaware of other examples where the analysis of quantum algorithmic performance yields similarly novel results.

Running the quantum algorithm on a quantum computer would efficiently find the actual cuts that achieve our lower bounds.  This provides an exponential speedup for finding these cuts, compared to known classical algorithms with rigorous guarantees.

\section{Acknowledgements}

This project was an outgrowth of a project which included Brandon Augustino, Madelyn  Cain, Swati Gupta, Eugene Tang and Katherine Van Kirk.
We are grateful to  them for getting us going.
We also thank Ojas Parekh and Kunal Marwaha for reading our manuscript.
We thank David Gamarnik for suggesting that we look at MIS.  We thank Joao Basso, Stephen Jordan and Mario Szegedy  for helpful comments and 
Leo Zhou for introducing the idea of the last paragraph of Section \ref{sec:methods}.
DR acknowledges support by the Simons Foundation under grant 376205. 

\bibliographystyle{unsrt}
\bibliography{main}

\end{document}